\documentclass[a4paper,
biblatex,     
]{jacow}
\usepackage{pdfpages,multirow,ragged2e} %
\makeatletter%
\ifboolexpr{bool{xetex}}
{\renewcommand{\Gin@extensions}{.pdf,%
		.png,.jpg,.bmp,.pict,.tif,.psd,.mac,.sga,.tga,.gif,%
		.eps,.ps,%
}}{}
\makeatother

\ifboolexpr{bool{xetex} or bool{luatex}} 
{}                                      
{\usepackage[utf8]{inputenc}}           

\usepackage[USenglish]{babel}

\begin{document}
	
	\title{resonant filling 
		for long-pulse operation of EuXFEL
		using iterative learning based feed forward}
	
	\author{B. Richter\thanks{bozo.richter@desy.de}\textsuperscript{,1}, J. Branlard, J. Einstein-Curtis\\
		Deutsches Elektronen-Synchrotron DESY, Hamburg, Germany \\
		\textsuperscript{1}also at Technische Universität Hamburg TUHH, Hamburg, Germany}
	
	\maketitle

	\begin{abstract}
		A future operation scenario of the European XFEL (EuXFEL) is long-pulse (LP) mode, where the 800 superconducting radio frequency (SRF) cavities of the linac are energized with RF pulse lengths of up to \SI{500}{\milli\second}. Compared to the current, short-pulse (SP) mode with less than \SI{2}{\milli\second}, the dominant mechanical eigenmodes of the cavities at a few hundred Hertz pose a significant control challenge, as the RF pulse extends beyond a multiple of their periods. To improve energy efficiency of LP operation, cavities are operated at bandwidths in the range of tens of Hertz, a factor of 10 narrower than in SP. Due to peak power limitation of the proposed high power amplifiers, even the fill time surpasses mechanical mode periods. This paper presents experimental evaluation of two feed forward adaptation approaches to facilitate resonant-filling despite mechanical mode transients: In a single-cavity regulation setup, the RF drive phase is adjusted to match the Lorentz force detuning induced cavity phase shift. In vector-sum RF regulation systems, the incident phase cannot be adjusted to each cavity individually, so the concept is inverted by adapting mechanical cavity tuner drives to match each individual cavity resonance trajectory with a fixed common RF drive phase.
	\end{abstract}
	
	\section{Introduction}
	Since the beginning of operation in 2017, the European XFEL (EuXFEL) provides ultrabright hard X-ray laser light to enable research in matter and technology. The X-ray flashes are driven by a linac that accelerates a bunched electron beam to electron energies of up to \SI{17.3}{\giga e\volt}. With an RF pulse repetition rate of \SI{10}{\hertz}, 2700 electron bunches are accelerated within a \SI{650}{\micro\second} flattop, using 800 superconducting \SI{1.3}{\giga\hertz} TESLA cavities distributed across 25 RF stations. RF pulse filling was optimized to reduce peak power needs and minimize reflections in the switching point to flattop \cite{FillXFEL}.
	
	Future option studies for an upgrade of the EuXFEL foresee an increase of the RF duty cycle at a reduced repetition rate of 1 or \SI{2}{\hertz}. The resulting flattop of up to \SI{500}{\milli\second} extends the time horizon of X-ray bursts and enables accelerating an increased number of bunches per pulse. However, the flattop then exceeds the oscillation periods of the dominant mechanical cavity modes. Consequently, modal transients excited during filling persist into the flattop, which indicates the need for a more advanced feed forward approach than \cite{FillXFEL} and implementation of active resonance control \cite{LPXFEL}. To improve energy, space and cost efficiency, the cavity RF bandwidth will be lowered by an order of magnitude to the range of tens of Hertz, enabling usage of RF sources with significantly lower peak power. However, the sensitivity of the RF field to detuning increases with narrower bandwidths, substantiating the need for active mechanical disturbance compensation.
	
	An iterative learning control (ILC) \cite{ILC}  based approach for piezo feed forward optimization in the context of the EuXFEL high duty cycle upgrade is introduced and simulated in \cite{andreaSimulation}. This work presents experimental results of the proposed approach and extends the concept to RF phase adaptation in single-cavity operation. 
	
	\section{SRF CAVITY MODEL}
	The RF field contained in a cavity leads to deformation of the cavity walls. The resulting geometry change shifts the RF resonance frequency $f_{\mathrm{res}}$ of the cavity. The static Lorentz force detuning (LFD) 
	\begin{equation}\label{eqn:lfd}
		\Delta f_\mathrm{LFD} = k V_\mathrm{acc}^2
	\end{equation}
	depends on the cavity accelerating voltage amplitude $V_\mathrm{acc}$ and the LFD constant $k$. For EuXFEL type TESLA cavities, $k$  varies between $-1.2$ and \SI{-0.7}{\hertz\per\mega\volt^2}.
	
\begin{figure}
		\centering
		\includegraphics[width=0.9\linewidth]{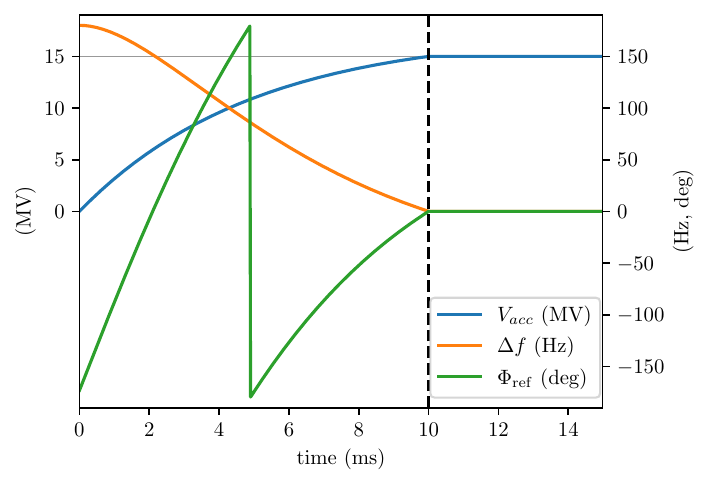}
		\caption{RF filling feed forward phase generation using cavity field $V_\mathrm{acc}$ with $f_{1/2}=\SI{32}{\hertz}$, the corresponding Lorentz force detuning $\Delta f$ with $k=\SI{-0.8}{\hertz\per\mega\volt^2}$, and the resulting phase trajectory $\Phi_\mathrm{ref}$. The RF drive amplitude is piecewise constant, with a step in the transition from filling to flattop which leads to non-smooth trajectories.}
		\label{fig:phase}
	\end{figure}

	Given an unenergized cavity with half bandwidth $f_{1/2}$, the trajectory of $V_\mathrm{acc}$ is given by
	\begin{equation}
		V_\mathrm{acc}(t) = 2a\sigma(t)\left(1-e^{-2\pi f_{1/2}t}\right)
	\end{equation}
	when exciting on resonance with the incident amplitude $a\sigma(t)$, where $\sigma(t)$ is the unit step function. The phase of the complex cavity voltage is described by
	\begin{equation}\label{eqn:phase}
		\Phi(t) = \Phi_0^{\ast}+2\pi\int_0^t\Delta f(\tau)d\tau.
	\end{equation}
	Assuming the detuning $\Delta f$ is the sum of an initial detuning $\Delta f_0$ and $\Delta f_\mathrm{LFD}$ from \eqref{eqn:lfd}, \eqref{eqn:phase} becomes 
	\begin{equation}\label{eqn:phase_expr}
		\dfrac{\Phi(t)}{2\pi} =  \dfrac{\Phi_0}{2\pi}+\Delta f_0 t + \dfrac{(2a)^2 k}{f_{1/2}} \left(f_{1/2}t + \dfrac{1}{\pi}e^{-\omega_{1/2}t} - \dfrac{1}{4\pi}e^{-2\omega_{1/2}t}\right)
	\end{equation}
	and constitutes the required phase modulation of the drive signal for resonant filling, see Fig.~\ref{fig:phase}.
	
	\section{ITERATIVE LEARNING CONTROL}
	Iterative learning control is a feed forward adaptation scheme that leverages knowledge obtained from preceding trials of a repeating reference tracking experiment.
	Here we follow \cite{andreaSimulation}, i.e. model-free single-input single-output ILC. Essentially, up to $m$ integral feedback controllers are instantiated, where $m$ is the number of output samples recorded during one repetition.  Each integral controller acts on the trajectory of values at one specific sampling index $k$ across subsequent trials $i$, in contrast to the classical application of integral controllers that act on subsequent samples $k$.
	
	The $n$ step-ahead proportional error-filtered ILC law
	\begin{equation}\label{eqn:ilc}
		u_{k}^{(i+1)} = u_{k}^{(i)} + \alpha H\Big[r_{k+n}^{(i)}-y_{k+n}^{(i)}\Big]
	\end{equation}
	penalizes the setpoint deviation $e=r-y$ at the sample index $k+n$ by adjusting the controller output $u$ at sample index $k$, where $i\in\mathbb{N}$ is the RF pulse number, $r$ is a tracking reference, $y$ is the system output, and $\alpha$ is the integrator gain or step size. $H$ is a filter applied to the whole error signal at trial $i$, blocking unwanted frequency components like measurement noise from entering the drive signal.
	
	\subsection{Phase ILC}
	For energy efficient cavity filling, the incident phase can be optimized to match the cavity accelerating field phase, thus exciting on resonance. A simplistic realization of this approach is shown in \cite{FillXFEL}, where the drive phase is a manually tuned exponential. On-resonance filling of a single cavity (SCAV) by an individual RF source is easiest to implement as a self excited (SEL) or phase-locked loop (PLL), where the high power RF drive phase is modulated by the measured cavity phase in a tight feedback loop. Nevertheless, iterative feed forward phase adaptation is investigated as a simple-to-implement alternative if an SEL controller is not available. 
	
	Assuming quasi-static LFD and neglecting mechanical resonance dynamics, a nominal feed forward phase $\Phi$ can be calculated from \eqref{eqn:phase_expr}. However, due to excitation of mechanical modes, amplifier nonlinearity, and inaccurate estimates of $k$ and $\omega_{1/2}$, the resulting phase trajectory will be suboptimal. Assuming the time constant of the actuator is less than the sampling time, a single-step-ahead ILC-like adaptation is employed. Unlike conventional ILC, the cavity phase serves as a time-varying tracking reference $r$. The learning algorithm minimizes the mismatch to the incident phase $u$. The RF drive phase can be either initialized by Eq. \eqref{eqn:phase_expr} or set to zero, if no estimates of $k$ and $\omega_{1/2}$ are available. To ensure the cavity is on resonance in the switching point from filling to flattop, detuning information is still required for static detuning adjustment.
	
	\subsection{Detuning ILC}
	The previously described phase adaptation leads to a system where LFD and mechanical cavity eigenmodes dictate the resulting phase and hence detuning trajectory, similar to a phase-locked loop. If a detuning trajectory shall be prescribed, e.g. after optimization to avoid excitation of mechanical resonances, the previous approach is not applicable. Additionally, if a single RF source drives multiple cavities in parallel (MCAV), an individual drive phase adaptation is not possible. In both scenarios, an iterative adaptation of detuning can be used to match the cavity to the RF drive.
	
	In MCAV, mean or median of $k$ and $\omega_{1/2}$ of all cavities can be used to generate an RF drive phase with Eq. \eqref{eqn:phase_expr}. 
	Then, the desired detuning trajectory is given by the time derivative of this drive phase trajectory. Estimating the cavity detuning using \cite{RichterLO}, ILC can be employed to reduce the detuning tracking error by acting on the mechanical piezo tuner.
	The delay of the piezo-detuning transfer function is accounted for by setting $n$ in Eq.~\eqref{eqn:ilc} accordingly, \cite{andreaSimulation}. The remaining mechanical dynamics are addressed by iterative adaptation. 
	
	\section{EXPERIMENTAL RESULTS}
	Experimental testing of the ILC algorithm was performed at the cryomodule test bench (CMTB). The test facility allows testing one EuXFEL-type cryomodule without beam. A CW-capable \SI{40}{\kilo\watt} inductive output-tube (IOT) supplies all 8 cavities of the installed cryomodule via a waveguide distribution system. Each cavity receives a different, fixed fraction of the total drive power. All tests are performed at a repetition rate of \SI{1}{\hertz}. 
	
	In pulsed mode, the IOT is subject to thermal effects leading to significant nonlinear behavior. In the phase ILC experiment, an additional ILC algorithm compensates the amplifier nonlinearity, whereas static inversion is used in the detuning ILC experiments. In both experiments, the error filter $H$ is a second order Butterworth backward-forward lowpass filter as described in \cite{andreaSimulation}.
	
	\subsection{Phase ILC}
	The phase ILC approach is tested on cavities of XM50.1, a cryomodule modified to provide half bandwidths as low as $f_{1/2}=\SI{10}{\hertz}$. Experiments are performed at $f_{1/2}\approx\SI{16}{\hertz}$ ($Q_L=4\times10^7$) and with a flat initial drive phase. The bandwidth of $H$ is set to \SI{4}{\kilo\hertz} and the ILC adaptation rate to $\alpha=0.2$. The detuning offset $\Delta f_0$ is adapted manually.
	
	\begin{figure}
		\centering
		\includegraphics[width=0.8\linewidth]{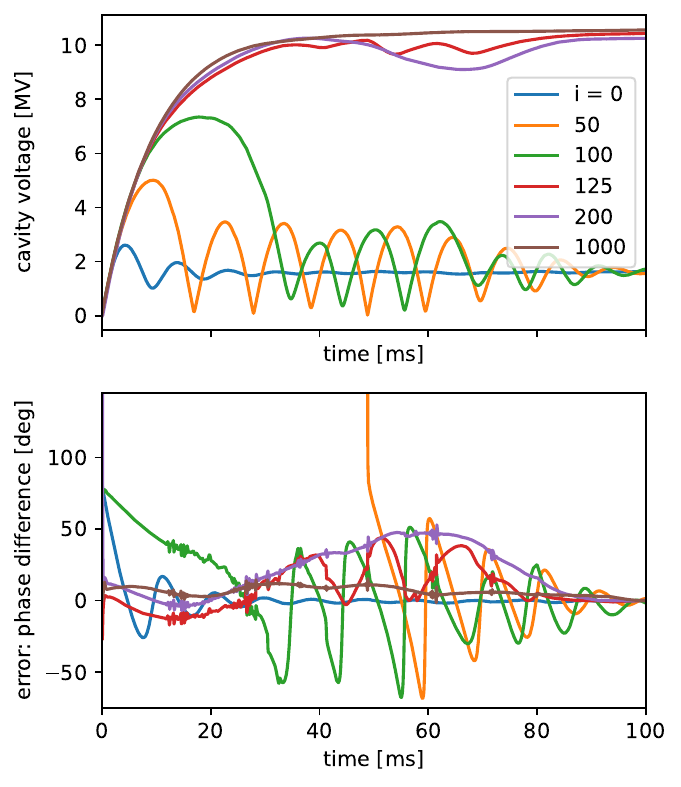}
		\caption{Voltages and phase errors in phase ILC after initialization with a flat RF drive phase.}
		\label{fig:phaseILC}
	\end{figure}
	
	After 125 iterations, the algorithm overcomes the ponderomotive drop, see Fig.~\ref{fig:phaseILC}. Nevertheless, substantial phase tracking errors of more than \SI{20}{\degree} remain even after 1000 trials. Additionally, notable discontinuities appear in the drive phase.
	
	Depending on the severity of microphonics, overcoming the ponderomotive drop is possible by adequate initialization of the drive phase through Eq. \eqref{eqn:phase_expr}. Then, convergence is faster, as phase trajectories learned before the drop do not need to be unlearned. For further improvement, the adaptation rate $\alpha$ can be optimized.
	
	\subsection{Detuning ILC}
	The detuning ILC approach was tested on 7 cavities of XM8 due to defective tuner cabling on the first cavity. Each cavity is set to its minimum half bandwidth $\SI{31}{\hertz} < f_{1/2} < \SI{50}{\hertz}$, i.e. $Q_L$ varying between $2.1$ and $1.3\times 10^7$. The piezo drive system provides feed forward tables of \SI{40}{\milli\second}, therefore the learning time horizon is limited. The tables are initialized with zeros instead of a model-based estimate like \cite{andreaSimulation}. The ILC delay $n$ is tuned experimentally. The bandwidth of $H$ is set to  \SI{300}{\hertz}.
	
	The detuning setpoint is generated using a desired average cavity voltage of \SI{10}{\mega\volt}, the cavity half bandwidth mean of $f_{1/2}=\SI{39.9}{\hertz}$, and $k=\SI{-0.9}{\hertz\per\mega\volt^2}$. Cavities are tuned to $\Delta f_0=+\SI{90}{\hertz}$ to minimize the error at $k=0$. 
	
\begin{figure}
		\centering
		\includegraphics[width=.9\linewidth]{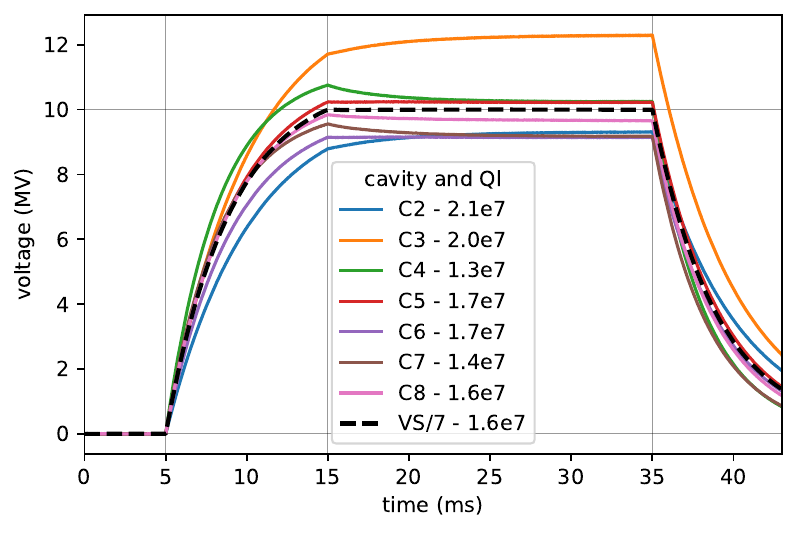}
		\caption{Cavity voltages after 100 pulses of detuning ILC.}
		\label{fig:detuningGrads}
	\end{figure}

\begin{figure}
		\centering
		\includegraphics[width=.9\linewidth]{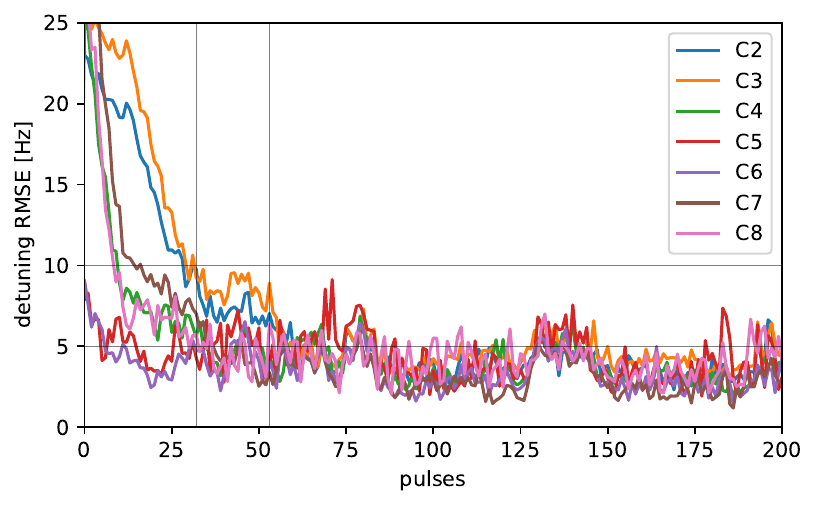}
		\caption{Evolution of experimental detuning RMSE (root-mean-square error) with 7 cavities under detuning ILC. Vector-sum RF control is enabled at pulse 53.}
		\label{fig:detuningRMSE}
	\end{figure}

	Figure \ref{fig:detuningGrads} shows how the spread in $f_{1/2}$ and waveguide power distribution fractions lead to different absolute voltages and relative fill levels per cavity during filling and flattop. Nevertheless, the vector sum is flat as desired. In Fig. \ref{fig:detuningRMSE}, the detuning RMSE shows a decay in all cavities below \SI{10}{\hertz} after 30 pulses. After enabling vector-sum RF feedback at pulse 53, RMSEs level out between 3 and \SI{6}{\hertz}. 
	
	\section{CONCLUSION}
	Both phase and detuning ILC were demonstrated experimentally. In single-cavity operation, phase ILC overcame the ponderomotive drop, but convergence was slow, with substantial residual phase errors. Further evaluation with non-flat phase initialization is required, though SEL remains preferable for on-resonance filling. In multi-cavity operation, detuning ILC reduced the tracking RMSE below 10 Hz for all seven cavities, demonstrating its feasibility for vector-sum RF regulation.
	
	Future investigations will focus on detuning ILC with extended learning horizons, narrower cavity bandwidths, higher cavity voltages, and detuning trajectory optimization to minimize excitation of mechanical modes. Further topics include model-based amplifier linearization, LFD characterization, adaptive piezo delay compensation, learning boundaries, and optimized error filtering. 
	
	\section{ACKNOWLEDGEMENTS}
	This work was funded in the context of the R\&D program of the European XFEL.
	This project has received funding from the European Unions Horizon Europe research and innovation program under grant agreement N.~101131435.

	\printbibliography
	
\end{document}